\documentstyle[aps,axodraw,epsfig,floats,preprint]{revtex}
\newcommand{\be}{\begin{equation}}
\newcommand{\ee}{\end{equation}}
\newcommand{\bea}{\begin{eqnarray}}
\newcommand{\eea}{\end{eqnarray}}

\begin{document}

\title{Signatures of HyperCharge Axions in Colliders}

\author{Ram Brustein, David H. Oaknin}
\address{Department of Physics,
Ben-Gurion University, Beer-Sheva 84105, Israel\\
 email: ramyb, doaknin @bgumail.bgu.ac.il}

\maketitle
\begin{abstract}
If  in addition to the standard model fields,  a new pseudoscalar
field that couples to
hypercharge topological number density, the hypercharge axion, exists,
it can be produced  in
colliders in association with photons or Z bosons, and detected by
looking for its decay into photons or Z's. For a
range of masses below a  TeV and coupling above a fraction of
1/TeV, existing data from LEP II and the Tevatron can already
put interesting  constraints, and in future colliders accessible
detection range is increased significantly. The hypercharge axion
can help in explaining
the matter-antimatter asymmetry in the universe.
 \end{abstract}

 If  in addition to the standard model fields,  a new pseudoscalar
field that couples to hypercharge topological number density, the
hypercharge axion (HCA), exists, it can be produced in colliders
in association with photons or Z bosons. Since the hypercharge
photon is a linear combination of the ordinary photon and the $Z$,
HCA couples to photons and $Z$'s and therefore it can be produced
in interactions involving photons or $Z$'s, and it can be
detected by looking for its decay into photons or $Z$'s. We show
that HCA can be produced in colliders in sufficient numbers to be
detected if its mass is below a few TeV and its coupling above a
fraction of TeV${}^{-1}$. In cosmology, HCA can exponentially
amplify hypercharge fields in the symmetric phase of the
electroweak plasma, while coherently rolling or oscillating
\cite{bo,Guendelman}, leading to the formation of a
time-dependent condensate of topological number density. This
condensate can be converted at the electroweak phase
transition \cite{gs}, under certain conditions, into baryons in
sufficient quantity to explain the observed baryon asymmetry in
the universe\cite{bo}.

Pseudoscalar fields with the proposed axion-like coupling appear
in several possible extensions of the Standard Model (SM)
\cite{dine}. They typically have only perturbative derivative
interactions and therefore vanishing potential, and acquire a mass
$m$, which could be as low as a fraction of an eV, or as high as
$10^{12}$ GeV, through non-perturbative interactions. The scale of
mass generation $F$, could be as high as the Planck scale but
could also be much lower, even down to the TeV range, however,
typically $m\ll F$. A particularly interesting mass range is  the
TeV range, expected to appear if mass generation is associated
with supersymmetry  breaking, and if HCA plays a role in
baryogenesis \cite{bo}. We focus on a singlet elementary HCA
whose only coupling to SM fields is to hypercharge fields, but
composite or non-singlet fields may also appear in some models
\cite{dine}.

The singlet HCA field $X$ has mass $m_X$, and the following
dimension 5 interaction Lagrangian density,
 \be \label{intlag}
{\cal L}_{X YY}= \frac{1}{8 M}\ X \epsilon^{\mu\nu\rho\sigma}
 Y_{\mu\nu} Y_{\rho\sigma},
 \ee
where $Y_{\mu\nu}$ is the $U(1)_Y$ hypercharge field strength, and the
coupling $\frac{1}{M}$ has units of mass${}^{-1}$. For QCD axions,
$M \sim F$, $F$ being the ``Peccei-Quinn" mass generation scale.
But in general, it is not always the case, and we
will therefore take $M$ to be a free parameter, and in particular
allow $M<F$, keeping the characteristic property $m_X\ll M$.
We will be interested in a range $M<100$ TeV,  and
$m_X<10$ TeV. Our model is therefore a two parameter model, and the
goal of our subsequent analysis is to determine for which domain in
$(m_X,M)$ space, HCA can be produced and detected in colliders.

The interaction (\ref{intlag}) is not renormalizable, and therefore
our model should be considered as an effective field theory, with
a cutoff. We may estimate the magnitude of radiative corrections
to SM quantities, such as the $\rho$ parameter,  to check that our
model is consistent with existing electroweak measurements, which
require additional corrections to be smaller than a fraction of a
percent. The relative magnitude of a loop with external gauge
bosons and an internal $X$ particle, compared to the same process
at tree level is approximately of order $\frac{1}{16 \pi^2}
\frac{m_X^2}{M^2}$,  assuming that $X$ is the heaviest particle
running in the loop \cite{Pich:1998xt}.  However, according to our
assumptions $m_X\ll M$, and therefore expected radiative
corrections are small. We will not attempt here a more detailed treatment
of radiative corrections.

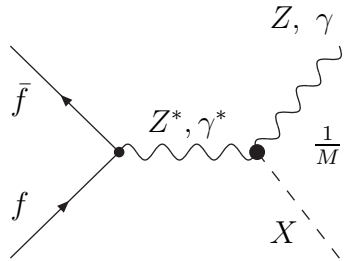
\begin{figure}
\begin{center}
\begin{picture}(75,120)(0,0)
 \ArrowLine(-20,20)(21,60)    \Vertex(21,60){2}
 \ArrowLine(21,60)(-20,100)
 \Text(-20,80)[l]{$\bar f$}
 \Photon(21,60)(73,60){3}{4} \Vertex(73,60){3}
 \Text(47,68)[b]{$Z^*,\gamma^*$}
 \Text(100,60)[b]{$\frac{1}{M}$} \Photon(73,60)(104,100){3}{4}
 \Text(-20,40)[l]{$f$} \Text(78,110)[l]{$Z,\ \gamma$}
 \DashLine(73,60)(104,20){5} \Text(78,30)[l]{$X$}
\end{picture}
\caption{\label{diag}
 Associated production of HCA in colliders.
The fermions $f$ can be either charged leptons or quarks.}
\end{center}
\end{figure}

\begin{figure}
\vspace{-.5in}
\centerline{\psfig{figure=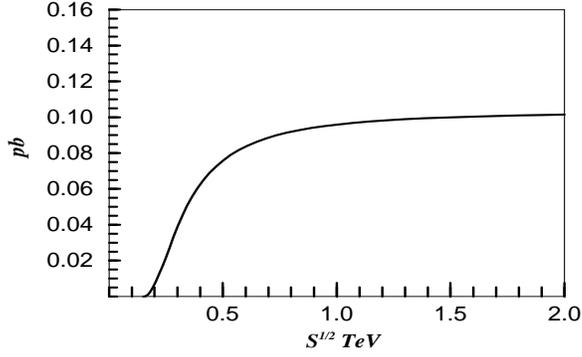,height=2.0in,width=3.2in}}
 \caption{\label{epluseminuscross}
 Total HCA associated production cross section in $e^+e^-$ colliders, for
 $m_X=150$ GeV, $M=1$ TeV.}
\end{figure}

\begin{figure}
\vspace{-.1in}
\centerline{\psfig{figure=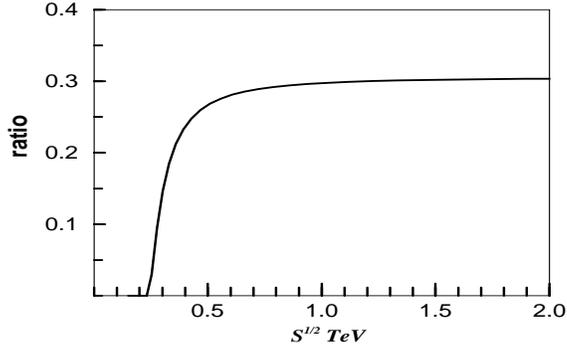,height=2.0in,width=3.2in}}
\caption{\label{ratiocross}
 {Ratio of cross section for associated production of HCA
with a $Z$ boson, to  cross section for associated production of HCA
 with a photon in $e^+e^-$ colliders, for $m_X=150$ GeV.} }
\end{figure}

 HCA  can be produced in colliders, if the energy is high enough, in
association with a photon or a $Z$. Recall that the hypercharge
photon $Y_\mu$ is a linear combination of the ordinary photon
$A_\mu$ and the Z boson $Z_\mu$,
 $Y_\mu=\cos\theta_W A_\mu-\sin\theta_W Z_\mu$,
where $\theta_W$ is the Weinberg angle. The  total
unpolarized cross sections for the two processes
 $f \overline{f} \rightarrow Z^{*},\gamma^{*} \rightarrow Z X$ and
$f\overline{f} \rightarrow Z^{*},\gamma^{*}\rightarrow \gamma X$,
(here $Z^*$ and $\gamma^*$ denote a virtual $Z$ and a virtual
photon, and $f$ is a charged fermion) can be computed from the
diagrams in Fig.~\ref{diag},
 \bea \label{eq:crosszx}
 && \sigma(f \overline{f} \rightarrow Z X)
= \frac{1}{N_c} \frac{\alpha}{48} \frac{1}{M^2} \frac{k^{3/2}}{s}
\frac{\sin^2 \theta_W}{\cos^2 \theta_W} \times
 \\ \nonumber &&
  \left(\left[\frac{\frac{1}{2}(c_v-c_a)}{s-M_z^2}\! +\!
\frac{\cos^2\theta_W}{s}\right]^2 \!+\!
\left[\frac{\frac{1}{2}(c_v+c_a)}{s-M_z^2}\! +\!
\frac{\cos^2\theta_W}{s}\right]^2 \right),
 \eea
 \bea
\label{eq:crossphx}
 && \sigma(f \overline{f} \rightarrow \gamma X)
  = \frac{1}{N_c} \frac{\alpha}{48} \frac{1}{M^2}
\frac{q^{3/2}}{s} \times
 \\ \nonumber &&
 \left(\left[\frac{\frac{1}{2}(c_v-c_a)}{s-M_z^2} +
 \frac{\cos^2\theta_W}{s}\right]^2 \!+\!
\left[\frac{\frac{1}{2}(c_v+c_a)}{s-M_z^2} +
 \frac{\cos^2\theta_W}{s}\right]^2 \right),
 \eea
where $k=(s - m_X^2 - m_Z^2)^2 - 4 m_X^2 m_Z^2$, $q=(s -m_X^2)^2$
and $\sqrt{s}$ is center of mass (CM) energy of the collision.  In
deriving eqs.(\ref{eq:crosszx},\ref{eq:crossphx}) we have assumed
that the fermions are effectively massless, $m_f/\sqrt{s} \ll 1$.
The parameters appearing in
eqs.(\ref{eq:crosszx},\ref{eq:crossphx}) are given in Table
\ref{tb1}: $c_v$ and $c_a$ are the vector and axial coupling of
the fermion to the Z, $N_c$ is the number of colors of the
fermion,  and takes into account averaging over initial colors.
Notice that for small $s$ there are kinematical thresholds for
both processes, $s>m_X^2$, $s>(m_X+m_Z)^2$  to allow
associated production with a photon, and with a Z, respectively.
 \begin{center}
    \begin{table}
    \begin{tabular}{||c||c|c|c||}
    & $ e $ & Up type quarks &  Down type quarks   \\
    \hline\hline
    $ c_v$ & $ -\frac{1}{2} + 2 \sin^2\theta_W   $ &
     $\frac{1}{2} - \frac{4}{3} \sin^2\theta_W   $ &
     $ -\frac{1}{2} + \frac{2}{3} \sin^2\theta_W   $  \\
     &$\simeq -0.03$ & $\simeq 0.19$ & $\simeq -0.34$ \\
    \hline
    $c_a$ & $-\frac{1}{2}$ & $ \frac{1}{2} $ & $- \frac{1}{2} $ \\
    \hline
    $N_c$ & $ 1 $ & $ 3$ & $ 3 $ \\
  \end{tabular}
  \caption{Parameter values for different fermions.}
  \label{tb1}
  \end{table}
   \end{center}
In Fig.~\ref{epluseminuscross} we plot the total cross section for
associated production of HCA in $e^+e^-$ colliders as a function
of CM energy $\sqrt{s}$,  for $m_X=150$ GeV, and $M=1$ TeV. The
cross section scales as $M^{-2}$, and therefore its magnitude for
different values of $M$ can be read off
Fig.~\ref{epluseminuscross}. Since
$q^{3/2}, k^{3/2} \sim s^3$ for large $s$, both cross sections
approach asymptotically a constant independent of the mass $m_X$
\cite{alphas},
 $\sigma(f \overline{f} \rightarrow Z X)\rightarrow
 \frac{1}{N_c} \frac{\alpha}{48} \frac{1}{M^2} \tan^2 \theta_W S_{f}$
 and
 $\sigma(f \overline{f} \rightarrow \gamma X)\rightarrow
 \frac{1}{N_c} \frac{\alpha}{48} \frac{1}{M^2} S_f$, where
$S_f$ is a numerical factor of order unity depending on the type
of  fermion. The rise towards the asymptotic value is governed by
the ratios ${m^2_X}/{s}$, and ${m^2_Z}/{s}$. Asymptotically, for
large $s$,
 $
\frac{ \sigma(f \overline{f} \rightarrow Z X)} {\sigma(f
\overline{f} \rightarrow \gamma X)}\simeq \tan^2\theta_W
 \simeq 0.3.
 $
The ratio of associated production cross section  with a Z to
 associated production cross section with a photon in $e^+e^-$ colliders
is shown in Fig.~\ref{ratiocross} for $m_X=150$ GeV. The ratio is,
of course, independent of $M$.

\begin{figure}
\vspace{-.5in}
\centerline{\psfig{figure=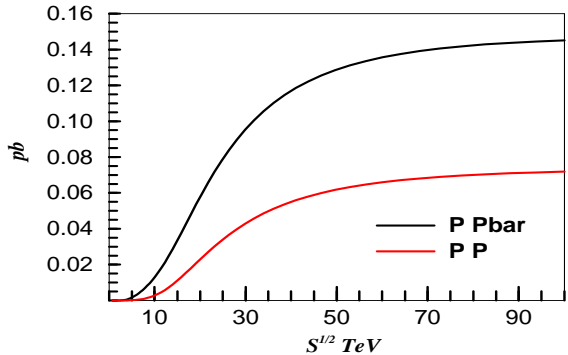,height=2.0in,width=3.2in}}
 \caption{\label{ppcross}
 { Total HCA associated production cross section
 in  hadronic colliders, for $m_X=800$ GeV, $M=1$ TeV.} }
\end{figure}

\begin{figure}
 \vspace{-.2in}
\centerline{\psfig{figure=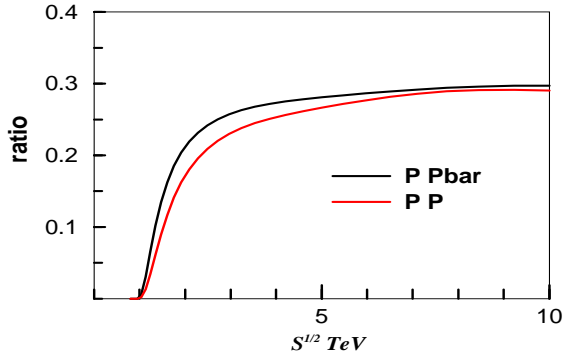,height=2.0in,width=3.2in}}
 \caption{\label{ppratio}
 {Ratio of cross section for associated production of HCA
with a $Z$ boson, to  cross section for associated production of HCA
 with a photon  in hadronic colliders, for $m_X=800$ GeV.} }
\end{figure}

To evaluate the cross section in hadron colliders such as the
Tevatron and LHC we follow the standard procedure of calculating
hadronic cross section $\sigma_{hadrons}$, for hadron collisions
at CM energy $\sqrt{s}$, from partonic ones ${\widehat
\sigma}_{ij}$ \cite{Eichten,pdg},
 $
 \sigma_{hadrons}=\int \sum\limits_{ij} f^{(a)}_i(x_1,Q^2)
 f^{(b)}_j(x_2,Q^2){\widehat \sigma}_{ij}(\sqrt{Q^2})dx_1 dx_2,
 $
where $f^{(a)}_i$ and $f^{(b)}_j$ are the partonic structure
functions of type-$i$ and type-$j$ quarks in type-$(a)$ and
type-$(b)$ hadrons, respectively, $x_1,x_2$ are momentum fractions
carried by partons 1 and 2, and $Q^2=x_1 x_2 s$ is a typical
momentum transfer in a partonic collision. For our process the
only possible contributions are from the 6 same-flavour
quark-antiquark partonic collisions, whose cross section is given
in eqs.(\ref{eq:crosszx},\ref{eq:crossphx}). Typically, the
hadronic cross section for a $P\bar{P}$ or a $PP$ machine is
similar to an $e^+e^-$ machine at about 0.1 CM energy. The
hadronic cross section of a $PP$ machine is suppressed with
respect to the cross section in a $P\bar P$ machine by an
additional factor associated with light antiquarks density within
the proton. In Fig.~\ref{ppcross} we show the results of
numerically evaluating total hadronic cross section for associated
production of HCA of mass $m_X=800$ GeV, for $M=1$ TeV. The ratio
of associated production cross section  with a Z to
 associated production cross section with a photon in $P\bar{P}$
 and $PP$ collisions,
 for the same parameters is shown in Fig.~\ref{ppratio}.
As previously noted, the rise towards the asymptotic value is
governed by the ratios ${m^2_X}/{\widetilde{s}}$, and
${m^2_Z}/{\widetilde{s}}$, where
$\sqrt{\widetilde{s}} \sim 0.1 \sqrt{s}$ is the typical invariant
mass of the partonic collisions. For small $\widetilde{s}$ there
are kinematical thresholds for both processes, $\widetilde{s}>m_X^2$,
 $\widetilde{s}>(m_X+m_Z)^2$,
to allow associated production
with a photon, and  with a Z, respectively. The
cross section scales as $M^{-2}$, and therefore its magnitude for
different values of $M$ can be read off Fig.~\ref{ppcross}.

\begin{figure}
\vspace{-.1in}
\centerline{\psfig{figure=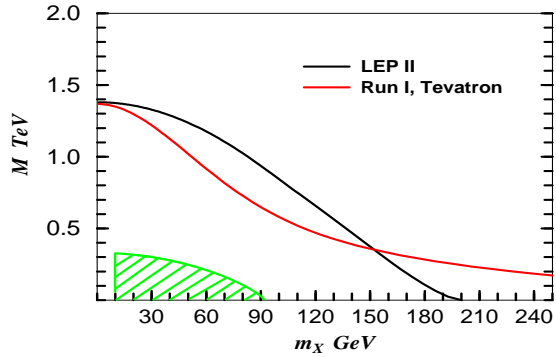,height=2.0in,width=3.2in}}
\caption{\label{leptev} {Expected number of HCA's in LEP II, and
Run I of the Tevatron. The curves correspond to 10 expected
events, below (above) the curve more (less) than 10 events are
expected. The shaded region is experimentally excluded.}  }
\end{figure}

\begin{figure}
\vspace{-.3in}
\centerline{\psfig{figure=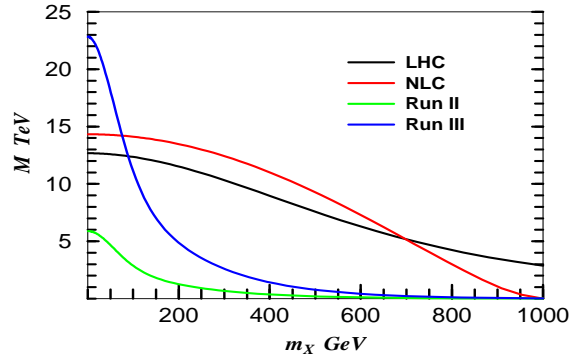,height=2.0in,width=3.2in} }
\caption{{ Expected number of  HCA's in future colliders. The
curves correspond to 10 expected events, below (above) the curve
more (less) than 10 events are expected.}} \label{future}
\end{figure}

Finally, we present the estimated number of produced HCA's in
colliders. The total number of HCA produced $N_{HCA}$, is given by
 $
 N_{HCA}={\cal L}\cdot \left(\frac{TeV}{M}\right)^2 \cdot
 \sigma\left(M=1\ TeV;s;m_X\right),
 $
where ${\cal L}$ is the total integrated luminosity available. If
we fix a number of events as the minimal number $N^{min}$ required
for detection of HCA, we obtain a region in $(m_X,M)$ space,
limited by the curve
 \be
\frac{M}{TeV}=\sqrt{ \frac{{\cal L}}{N^{min}}}
 \sqrt{\sigma\left(M=1\ TeV;s;m_X\right)},
\label{ngraph}
 \ee
shown in Fig.~\ref{leptev} for $N^{min}=10$ events for  LEP II and
for Run I of the Tevatron collider.  Existing experimental
searches \cite{opal1},  have set upper bounds only on the cross
section $\sigma(e^+e^-\rightarrow XZ)$, of the order of $0.1 pb$,
for $m_X<90$ GeV, assuming that $X\rightarrow \gamma\gamma$ (see
later). In Fig.~\ref{leptev}, we have shown the excluded region in
$(m_X,M)$ space, assuming a uniform upper bound of $0.1 pb$, for
$10 GeV< m_X< 90 GeV$. We expect a better reach by a specialized
analysis adapted to our model. At the Tevatron some analysis was
performed \cite{cdfd0} but the results are presented in a way that
is not directly useful. The reach of future colliders is shown in
Fig.~\ref{future}, where we have fixed $N^{min}=10$ events. For
the graphs in Figs.~\ref{leptev}, \ref{future}, we have chosen
some generic parameters characterizing the various machines,
summarized in Table \ref{tb2}, but since graphs scale with
parameters as in eq.(\ref{ngraph}), it is easy to adapt them to
different parameters. In the region below the curve more than 10
events are expected with luminosities given in Table \ref{tb2},
while in the region above the curve less than 10 events are
expected. The vertical position of the curve scales as the square
root of the integrated luminosity. Our choice of $N^{min}=10$
events is motivated by our assumption that the SM background for
our process is small (see below). Since curves in
Figs.~\ref{leptev}, \ref{future}, scale as $1/\sqrt{N^{min}}$, it
is easy to determine the reach for different values of $N^{min}$.
\begin{center}
    \begin{table}
    \begin{tabular}{||c||c||c||c||}
   collider& type & integrated luminosity & $\sqrt{s}$   \\
    \hline\hline
      LEP II &$ e^+e^-$&$200 pb^{-1} $&   200  GeV   \\
    \hline
      NLC  & $ e^+e^-$ & $20 fb^{-1} $ &    1  TeV   \\
    \hline
      &   & Run I: $100 pb^{-1} $ &    1.8  TeV    \\
      Tevatron  & $ P  \bar P$  & Run II:  $2 fb^{-1} $ &    2  TeV    \\
    &   & Run III:  $30 fb^{-1} $ &    2  TeV    \\
    \hline
      LHC  & $ P  P$ & $20 fb^{-1} $ &   14\ TeV   \\
  \end{tabular}
  \caption{Parameter values for different colliders.}
  \label{tb2}
  \end{table}
  \end{center}

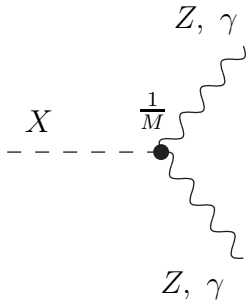
\begin{figure}
\begin{center}
\begin{picture}(75,120)(0,0)
 \DashLine(15,60)(73,60){5} \Vertex(73,60){3}
 \Text(27,68)[b]{$X$}
 \Text(70,73)[b]{$\frac{1}{M}$} \Photon(73,60)(104,100){3}{4}
 \Text(78,110)[l]{$Z,\ \gamma$}
 \Photon(73,60)(104,20){3}{4} \Text(73,10)[l]{$Z,\ \gamma$}
\end{picture}
{\caption  {Decay of HCA.}
\label{lifetime} }
\end{center}
\end{figure}

The produced HCA's decay into photons or $Z$'s in rates that
can be evaluated from the diagrams in
Fig.~\ref{lifetime}. If $m_X<m_Z$, HCA can only decay into two
photons, with a rate $\Gamma$ given by
$
\Gamma(X \rightarrow 2 \gamma) = \frac{1}{32\pi} \cos^4 \theta_W
\frac{m_X^3}{M^2}.
$
If   $m_Z<m_X$, HCA has an additional decay channel,
$
 \Gamma(X \rightarrow Z \gamma) = \frac{1}{32\pi}
 \cos^2\theta_W \sin^2 \theta_W
 \left(\frac{(m_X^2-m_Z^2)^3}{m_X^3} \frac{1}{M^2}\right).
$
Finally, if $m_X>2 m_Z$, HCA can also decay into two $Z$'s,
$
 \Gamma(X \rightarrow 2 Z) = \frac{1}{32\pi}
  \sin^4 \theta_W \frac{1}{M^2}(m_X^2 - 4 m_Z^2)^{3/2}.
$
The total decay rate is the sum of  rates for all the possible
decay channels. The signature of a decay of HCA is therefore,
depending on the mass $m_X$, a diphoton,  photon Z or a ZZ systems
in definite ratios.

The actual reach of colliders will be determined by the background
to our process.  In the SM there are no tree level processes
leading to a final state with three neutral gauge bosons, which is
the signature of our process. But there are one loop processes,
which do lead to a  final state with three neutral gauge bosons
\cite{jikia}. However, they do not seem to  provide substantial
background for values of $M$ below 50 TeV. We will not attempt
here a more detailed treatment of backgrounds.

If NLC can be operated as a photon collider, and its CM energy can
be tuned, then HCA can be produced directly, through the reverse
process of the one shown in  the diagram in Fig.~\ref{lifetime}.
The cross section for resonant production of $X$ in photon
colliders $\sigma= \frac{\pi}{4} \cos^4 \theta_W \frac{1}{M^2}\sim
150 \left(\frac{TeV}{M}\right)^2 pb$, is substantially larger than
associated production cross section in $e^+e^-$ colliders, and we
believe that this option should be investigated further.

Our conclusions are that existing data from LEP II and Run I of
the Tevatron can be used to rule out a region in $(m_X,M)$
parameter space, as shown in Fig.~\ref{leptev}, however a specific
analysis, similar to the analysis performed in \cite{opal1}, but
with the particular details of our model, has to be carried out to
determine the exact region. In future colliders HCA can be
detected provided that its mass is not too high, and its coupling
is not too weak as shown in Fig.~\ref{future}. The signature of
HCA are events with 3 neutral gauge bosons ($\gamma$ or $Z$), such
as diphoton dijet events. Production of $Z$'s requires higher
energy, but once enough energy is available (see
Figs.~\ref{epluseminuscross},\ref{ppcross}), and if the mass $m_X$
is high enough, all possibilities will show up, with  definite
calculable reduction factors of  $0.3$ or smaller for each
additional $Z$. The measured ratios can serve as a powerful check
for verifying discovery. Higher effective luminosity increases the
reach in coupling $M$, while higher energy increases the reach in
mass $m_X$. In particular, for luminosities and energies given at
Table \ref{tb2}, LHC and NLC have a similar reach. In comparison,
Run III of the Tevatron has a better reach in $M$ for smaller
$m_X$, and worse reach in $m_X$ for smaller $M$.

HCA can play a role in baryogenesis only for a certain range of
parameters, in particular, being conservative $m_X < 10$ TeV,
while $M$ appears only in combination with the initial coherent
HCA amplitude, and is therefore not constrained directly.
Detection of HCA in colliders with a mass below  $10$ TeV will
support the hypothesis that HCA can help in explaining the
observed matter - antimatter asymmetry in the universe.

\acknowledgments
 We gratefully acknowledge stimulating and useful
discussions with G. Kane, and we thank  G. Eilam for discussions
about radiative corrections and photon colliders and M. Oreglia
for comments about \cite{opal1}. Work supported  by the Israeli
science Foundation. D.O. is supported  by the Ministry of
Education and Science of Spain.


\begin{references}



\bibitem{bo}
R.~Brustein and D.H.~Oaknin,
 Phys. Rev. Lett. {\bf 82} (1999) 2628;  hep-ph/9901242.

\bibitem{Guendelman}
E.I.~Guendelman and D.A.~Owen, Phys. Lett. {\bf B276} (1992) 108.

\bibitem{gs}
M.~Giovannini and M.E.~Shaposhnikov, Phys. Rev. Lett. {\bf 80}
(1998) 22;
C.~Thompson, Phys. Lett. {\bf B422} (1998) 61.


\bibitem{dine}
M.~Dine, P.~Huet, R.J.~Singleton and L.~Susskind,
Phys. Lett. {\bf B257} (1991) 351.

\bibitem{Pich:1998xt}
See, for example, A.~Pich, hep-ph/9806303.


\bibitem{alphas}
Apart from a logarithmic dependence of $\alpha(\sqrt{s})$.


\bibitem{Eichten}
E.~Eichten, I.~Hinchliffe, K.~Lane and C.~Quigg,
Rev. Mod. Phys. {\bf 56} (1984) 579.

\bibitem{pdg}
C.~Caso {\it et al.},
Eur. Phys. J. {\bf C3} (1998) 1.


\bibitem{opal1}
K.~Ackerstaff {\it et al.} [OPAL Collaboration],
Phys. Lett. {\bf B437}, 218 (1998).


\bibitem{cdfd0}
F.~Abe {\it et al.} [CDF Collaboration],
Phys. Rev. Lett. {\bf 81}, 1791 (1998);
B.~Abbott {\it et al.} [D0 Collaboration], Phys. Rev. Lett. {\bf
82} (1999) 2244;
P.J.~Wilson [CDF Collaboration],
preprint Fermilab-Conf-98/213-E.

\bibitem{jikia}
G.~Jikia,
Nucl. Phys. {\bf B405}, 24 (1993);
G.~Jikia and A.~Tkabladze,
Phys. Lett. {\bf B323}, 453 (1994); Phys. Lett. {\bf B332}, 441 (1994).

\end{references}
\end{document}